\documentclass[useAMS,usenatbib]{article}
\usepackage{graphicx}
\usepackage{amssymb}
\usepackage{amsmath}
\usepackage{times}
\usepackage{color}
\usepackage{enumerate}

\newcommand{\msun}{M_\odot}
\newcommand{\Ha}{\mathrm{H}\alpha}
\newcommand{\Hb}{\mathrm{H}\beta}

\begin{document}

\begin{center}
   
{\large \bf Supercritical Accretion Discs in Ultraluminous X-ray Sources and SS 433}
\end{center}

\noindent
Sergei Fabrika$^{1,2*}$, Yoshihiro Ueda$^3$, Alexander Vinokurov$^1$, Olga Sholukhova$^1$, Megumi Shidatsu$^3$  \\

\noindent
$^1$ Special Astrophysical Observatory, Russia \\
$^2$ Kazan Federal University, Russia \\
$^3$ Kyoto University, Japan

\vspace{1cm}

\noindent
The black hole mass and accretion rate in Ultraluminous X-ray sources
(ULXs) in external galaxies, whose X-ray luminosities exceed those of
the brightest black holes in our Galaxy by hundreds and thousands of
times$^{1,2}$, is an unsolved problem.
Here we report that all ULXs ever spectroscopically observed have about 
the same optical spectra apparently of WNL type (late nitrogen 
Wolf-Rayet stars) or LBV (luminous blue variables) in their hot state, 
which are very scarce stellar objects. We show that the spectra do 
not originate from WNL/LBV type donors but from very hot winds from 
the accretion discs with nearly normal hydrogen content, which have
similar physical conditions as the stellar winds from these stars.
The optical spectra are similar to that of SS 433, the only known 
supercritical accretor in our Galaxy$^{3}$, although the ULX spectra 
indicate a higher wind temperature. Our results suggest that 
ULXs with X-ray luminosities of $\sim 10^{40}$\,erg~s$^{-1}$
must constitute a homogeneous class of objects, which most likely 
have supercritical accretion discs.

\newpage

ULXs have luminosities exceeding the Eddington limit for a typical
stellar-mass black hole$^{1,2}$, $\sim 2 \times 10^{39}$ erg
s$^{-1}$. Despite their importance in understanding the origin of
supermassive black holes that reside in most of present galaxies,
several critical questions are still open. 
What black hole masses do the ULXs contain? Do they posses standard
accretion discs? Do they belong to a homogeneous class of objects? The
most popular models for the ULXs involve either intermediate mass black
holes (IMBH, $10^3$--$10^4 \msun$) with standard accretion discs$^{4}$ 
or stellar-mass black holes ($\sim 10 \msun$) accreting at super
Eddington rates$^{5}$. Except for transient ULXs, both scenarios require 
a donor with a high mass loss rate in a close binary 
(like a massive star and an AGB star), which means
a short evolution stage.

While the X-ray properties of the ULXs are astonishing, it has been
impossible, from the X-ray data alone, to distinguish even the main ULX
models proposed. Optical spectroscopy may provide us with unique 
information on the ULXs. Because the ULX optical counterparts are very 
faint targets ($\sim$\,22--24~mag), however, very deep observations 
with the largest telescopes are required.

Using the Japanese 8.2m telescope Subaru, we have obtained 
highest-quality optical spectra of four nearest, bona fide ULXs that have 
unambiguous optical counterparts, single star-like objects~--- 
Holmberg\,II\,X-1, Holmberg\,IX\,X-1, NGC\,4559\,X-7, and NGC\,5204\,X-1. 
All details on the observations and data reduction can be found in 
Supplementary Section~1.

Fig.\,1 shows the spectra of the ULX optical counterparts (the full
range data are given in Supplementary Fig.\,1). Main features in all the
spectra are the bright He\,II\,$\lambda 4686$, hydrogen H$\alpha$, and
H$\beta$ emission lines.  The lines are obviously broad; the widths
range from 500 to 1500\,km s$^{-1}$.  In some objects we detect broad
He\,I\,$\lambda 6678, 5876$ emission lines. The spectra are very blue
(Supplementary Fig.\,1), in agreement with the photometric results$^{6}$.

We reveal that the ULX spectra are similar to those of WNL stars, or extreme
hot Of supergiants --- transition stars, or those of LBVs in their
compact hot states$^{7-9}$.
All these are massive stars in neighboring 
evolutional stages. This indicates the presence of hot outflow in the 
binary system: it could be a stellar wind from the donor, an irradiated 
surface of the accretion disc, or a powerful disc wind.
The spectra are also similar to that of SS\,433$^{10}$, the only
known supercritical accretor in our Galaxy, a close binary consisting of
an A-type supergiant and a stellar-mass black hole. SS 433
apparently exhibits a WNL-type spectrum$^{3,11}$ because the physical 
conditions of its disc wind may be similar to those of
stellar winds from WNL stars. The impression of the spectral similarity
is strengthened because the absolute visual magnitudes of both WNLs
and SS\,433 are in the same range as in the ULX counterparts$^{6,12}$
(Fig.\,2). Note that these WNLs are not 
WN\,5-7 stars, whose spectra are well described
with a pure-helium models$^{12}$, but LBV-like WN\,9-11 stars, which
contain up to 70\,\% hydrogen.
Such spectra of high luminosities with prominent He II emission lines
have never been observed from any stellar-mass black hole X-ray binaries
except for SS 433 and those having WNL donors.

However, neither very hot WNL nor LBV examples
exhibit such strong He\,II\,$\lambda 4686$ emission lines relative to
the hydrogen lines. If the abundance of hydrogen in the ULX donors were
two times smaller from the Solar value, then it could make the He/H
ratio five times larger, as observed in these ULX spectra (Supplementary
Table 1).  This contradicts with the non-enhancement of the He\,I and
Pickering He\,II lines indicating nearly normal abundance of hydrogen.
Hence, the wind in the ULXs must be even hotter and more highly ionized
than stellar winds in WNL or LBV stars.

All the spectra of the ULXs we obtained are surprisingly similar to one
another. The averaged width (FWHM) of the He\,II line and of H$\alpha$
are $\approx$870~km s$^{-1}$ and $\approx$1000~km s$^{-1}$, respectively.
Strikingly, we find that almost all 
ULX counterparts ever spectroscopically observed exhibit 
spectra similar to those of WN\,9-11 stars, 
particularly in their broad and strong He\,II emission,
including NGC\,5408\,X-1, NGC\,1313\,X-2, M81 X-6, and M\,101\,ULX-1$^{15-18}$. 
In Fig.\,3 we plot the He\,II and H$\alpha$ line widths of our four
objects supplemented with NGC\,5408\,X-1, which has a simultaneous 
spectrum including both He\,II and H$\alpha$. 

We study the spectra of the ULX counterparts in the He\,II diagram$^{7}$,
where the relation between the line width and equivalent width of
the He\,II line is plotted (Supplementary Fig. 2). 
Here the line width represents the terminal velocity of a stellar wind,
while the equivalent width reflects its photosphere temperature and mass
loss rate. We also include hottest transition stars from O2If to 
WN7ha$^{9}$, recorded LBV transitions$^{8}$, and SS\,433$^{10}$. 
The ULXs and SS\,433 occupy a region of the hottest transition
stars O2If/WN5~--~O3.5If/WN7 stars$^{9}$. However, their behavior
in the He\,II diagram is nothing like stars. They exhibit night-to-night
variability both in the line width and equivalent width by a factor of
2--3. Variability in the radial velocity of the line is also
detected with amplitudes ranging from 100~km s$^{-1}$ in Holmberg\,IX to
350~km s$^{-1}$ in NGC\,5204 (Supplementary Fig. 3).

First, we can exclude the case where these ULXs actually have WNL donors 
and their stellar winds produce the observed optical spectra. Indeed,
the rapid variability of the He\,II line-width is difficult to be
explained, because the wind terminal velocity in stars is determined by
the surface gravity. A more critical problem is that a WNL star means
the wind-fed accretion regime, which is not effective$^{18}$. To
provide the observed X-ray luminosity, $L_X \sim 10^{40}$\,ers s$^{-1}$,
one needs an unrealistically powerful wind from the donor if the
accretor is a stellar mass black hole. If the primary star
were an intermediate mass black hole (IMBH), then the variability of the
donor's orbital velocity would be much greater than observed
(see Supplementary Section~4 for details). 
Another obvious argument is the high-quality optical spectrum of 
NGC\,7793\,P13 indicating a B9Ia supergiant companion$^{14}$, which itself
cannot produce the He\,II emission line. Therefore, the He\, II line
observed in NGC\,7793\,P13 must be formed in the second companion, 
namely, in a photoionized wind from the accretion disc 
(Supplementary Section~4). 

We next consider a possibility that the He\,II line is formed in the
standard accretion disc around a massive black hole.  To produce lines
in emission, a strong irradiation of the disc surface by the central
source is required. The He\,II lines trace hotter regions located closer
to the black hole than the hydrogen lines, and therefore the He\,II line
width is expected to be notably larger than that of H$\alpha$ in a
normal geometry. Indeed, broad He\,II and hydrogen emission lines from the
self-irradiated disc are observed in Galactic stellar-mass black hole
X-ray binaries in ourbursts, such as V404\,Cyg$^{19}$, GRO
J1655--40$^{20,21}$ and GX 339--4$^{22}$.  In all these cases, the He II
emission lines are exactly broader than the hydrogen ones.  These
sources did not reach the supercritical regime; the most reliable
determination of the distance toward V404\,Cyg based on the astrometric
VLBI observations indicates that the luminosity was not super-Eddington
during its famous 1989 outburst$^{23}$. We can interpret that their
irradiated discs are not blocked by disc winds completely, and hence we
observe the He\,II emission directly at the disc surface. By contrast,
the ULX spectra show that the He\,II line is narrower than the H$\alpha$
(Fig.\,3) line. It is impossible to explain this fact by irradiated
discs unless exotic models are invoked.

We finally examine if the optical spectra of our ULXs can be explained
by a supercritical accretion disc (SCAD) with a stellar-mass black hole.
Indeed, the only known supercritical accretor SS\,433 shows a similar 
optical spectrum to ours, which is produced by the disc wind from the 
SCAD$^3$. In SS\,433 the He II line is narrower than H$\alpha$\,$^{10,24}$. 
The same is valid in WNL stars and LBV in their hot state$^{9,25}$.
This suggests that the He\,II and H$\alpha$ lines are formed in
different parts of radiatively accelerated disc winds, where more 
ionized gas located closer to the source has smaller outflow velocites.

In Supplementary Table\,1, we present the mean equivalent widths 
(relative fluxes) of main emission lines in the ULX spectra with 
respect to SS\,433. While the line fluxes in all our ULXs are very similar, 
the indicated ionization degrees are higher than that of SS\,433; 
the hydrogen and He\,I lines are $\approx 12$ times weaker than those in
SS\,433, while the two observed He\,II lines are only two times
weaker. These testify less dense but hotter disc winds than that in SS\,433.

While the luminosity is proportional to the mass accretion 
rate in standard discs, it is expected to have a logarithmical 
dependence in SCADs$^{26,5}$ (Supplementary Section 5), because 
the excess gas is expelled 
as a disc wind and the accreted gas is advected with the photon 
trapping, contributing little to the photon luminosity. Instead, 
the SCAD luminosity has the same order of the Eddington luminosity, 
$L_{\rm Edd} \approx 1.5 \times 10^{39} m_{10}$\,erg s$^{-1}$, 
where $m_{10}$ is the black hole mass in units of 10 solar masses.
The mass accretion rate can provide a factor of several in the 
logarithmic term. Besides that, the funnel in the SCAD wind 
will collimate the X-ray radiation to an observer also with a
factor of a few when observed with an inclination angle smaller than
40--50 degrees$^{27}$. Thus, the apparent X-ray luminosity of an ULX with
a stellar mass black hole may well be up to $\sim 10^{41}$ erg s$^{-1}$. 
The SCADs are able to account for the huge X-ray luminosities of 
the ULXs. 

The UV and optical luminosity may strongly depend on the original 
mass accretion rate $\dot M_0$, because these budgets are mainly 
produced by the reprocess of the strong irradiation from the SCAD's 
wind (the excess gas). Using simple relations for the SCADs, we 
find that the optical luminosity of the wind $L_V \propto \dot M_0^{9/4}$ 
and the wind temperature $T \propto \dot M_0^{-3/4}$ (see Supplementary 
Section 5 for more detail discussion). Accordingly, we find that 
the mass accretion rates in the ULXs listed in Fig.\,2 may be by a 
factor of 1.5--6 smaller and their wind temperatures are by 1.4--4 
times higher than those in SS\,433. SCAD models can explain both the 
relative dimness of the ULXs in the optical band and higher ionization 
states of their disc winds than those of SS\,433. Thus, we can interpret 
that SS\,433 is intrinsically the same as ULXs but an extreme case 
with a particularly high mass accretion rate, which could explain the 
presence of its persistent jets.

{\bf Acknowledgements}~~

\noindent
The authors are grateful to V.\,Shimansky, D.\,Nogami for helpful
comments, and T.\,Hattori for his support in our observations. The
research was supported by the Russian RFBR grant 13-02-00885, the
Program for Leading Scientific Schools of Russia N\,2043.2014.2, 
the Russian Scientific Foundation (grant N\,14-50-00043) and the
JSPS KAKENHI Grant number 26400228. S.F. acknowledges support of the Russian 
Government Program of Competitive Growth of Kazan Federal University.
Based on data collected at Subaru Telescope, which is operated by the 
National Astronomical Observatory of Japan and on data obtained from 
the ESO Science Archive Facility (NGC\,5408\,X-1, ID 385.D-0782).

{\bf Author contributions}~~ S.F. and Y.U. proposed the observations and
wrote the paper with comments from all authors. The observations
were carried out by S.F., Y.U. and M.S. A.V. and O.S. prepared the
observational details and reduced the data. S.F., Y.U. and
A.V. discussed the results and overall science case with contribution
from the remaining authors.

{\bf Author contributions}~~ The authors declare no competing financial 
interests. All correspondence should be addresed to S. Fabrika 
(email: fabrika@sao.ru)

\clearpage

\noindent
{\bf Figure 1: Spectra of the ULX optical counterparts} from top to
bottom, Holmberg\,II, Holmberg\,IX, NGC\,4559, and NGC\,5204 in blue
({\bf a}) and red ({\bf b}) spectral regions. The spectra are normalized
for better inspection. The most strong are the He\,II $\lambda 4686$ 
line and the hydrogen lines H$\alpha$ $\lambda 6563$ and H$\beta$ $\lambda 4861$. 
The broad He\,I\,$\lambda 6678$ line is also detected. Narrow nebular emission
in H$\beta$ and [O\,III] $\lambda \lambda 4959, 5007$ lines is oversubstacted
in two bottom spectra. Although the hydrogen lines are contaminated with 
the nebular emission, their broad wings are clearly seen.  \\ \\

\noindent
{\bf Figure 2: Absolute magnitudes of all well-studied ULXs:}
Holmberg\,IX~X-1, NGC 5204~X-1, NGC\,4559~X-7, IC\,342~X-1,
NGC\,5408~X-1, M\,81~X-6, M\,101~ULX-1, Holmberg\,II~X-1,
NGC\,6946~ULX-1, NGC\,1313~X-1, X-2)$^{6}$ and SS\,433 (shadowed) with
some updates in the distances in the last four ULXs$^{13}$ and
NGC\,7793~P13$^{14}$. \\ \\

\noindent
{\bf Figure 3: Emission line widths of He\,II and H$\alpha$}
of Holmberg\,II~X-1, NGC\,4559~X-7, NGC\,5408\,X-1, Holmberg\,IX~X-1, 
and NGC\,5204~X-1, from left to right, all extracted from simultaneous 
spectra. Systematic errors are taken into account 
in the error bars of H$\alpha$ because of the nebula line substraction.
The error on the He\,II line-width of NGC\,5204~X-1 
is large because of its fainter flux. Note that the averaged He\,II line
width of NGC\,1313~X-2 is 580\,km/s$^{16}$ measured in the spectra 
without H$\alpha$.

\clearpage
\begin{figure*}[p]
\begin{center}
\hspace*{-2.2cm}
\includegraphics[width=16cm]{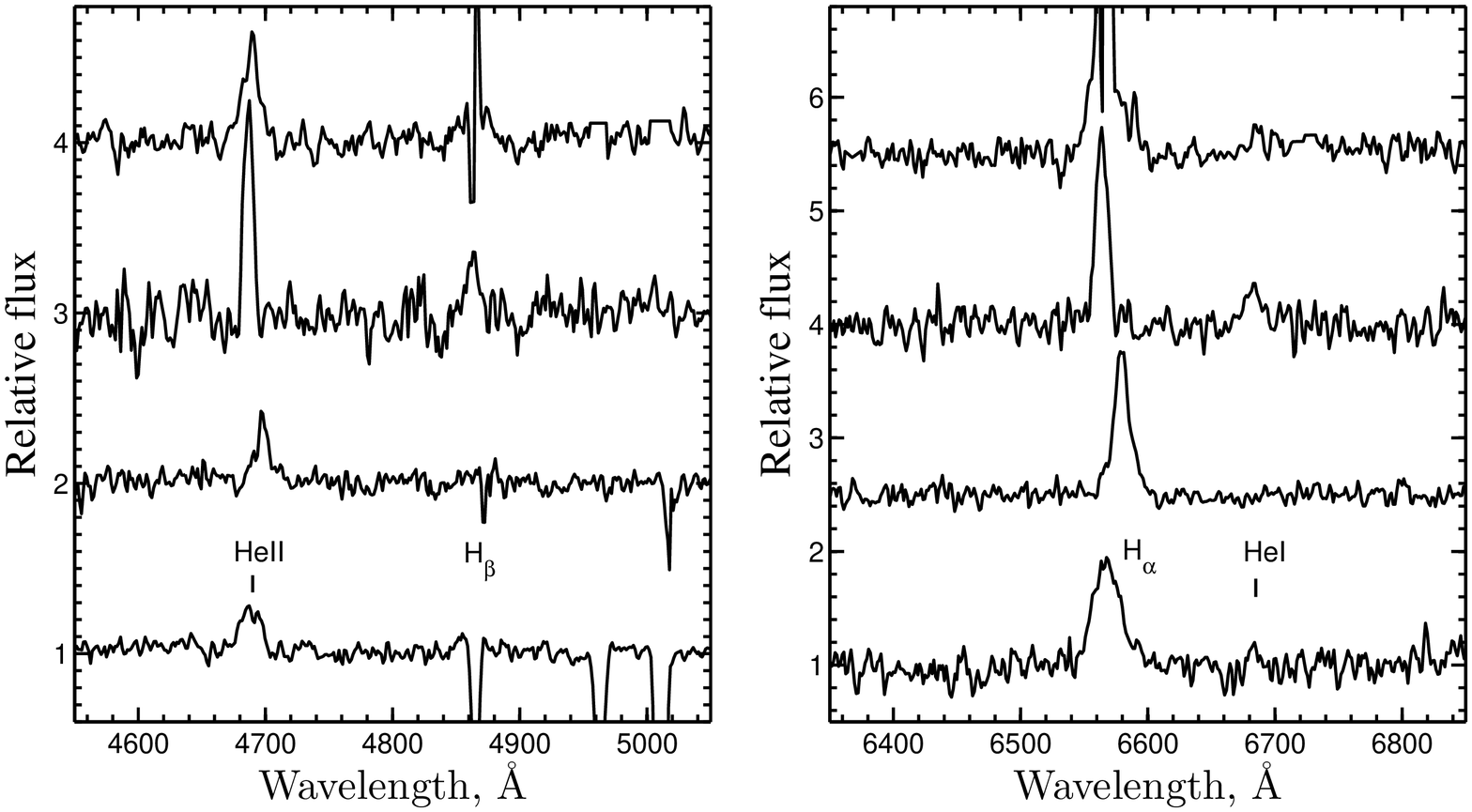}
\end{center}
\caption{} 
\label{fig1_spectra}
\end{figure*}

\clearpage
\begin{figure*}[p]
\begin{center}
\hspace*{-0.8cm}
\includegraphics[width=11cm]{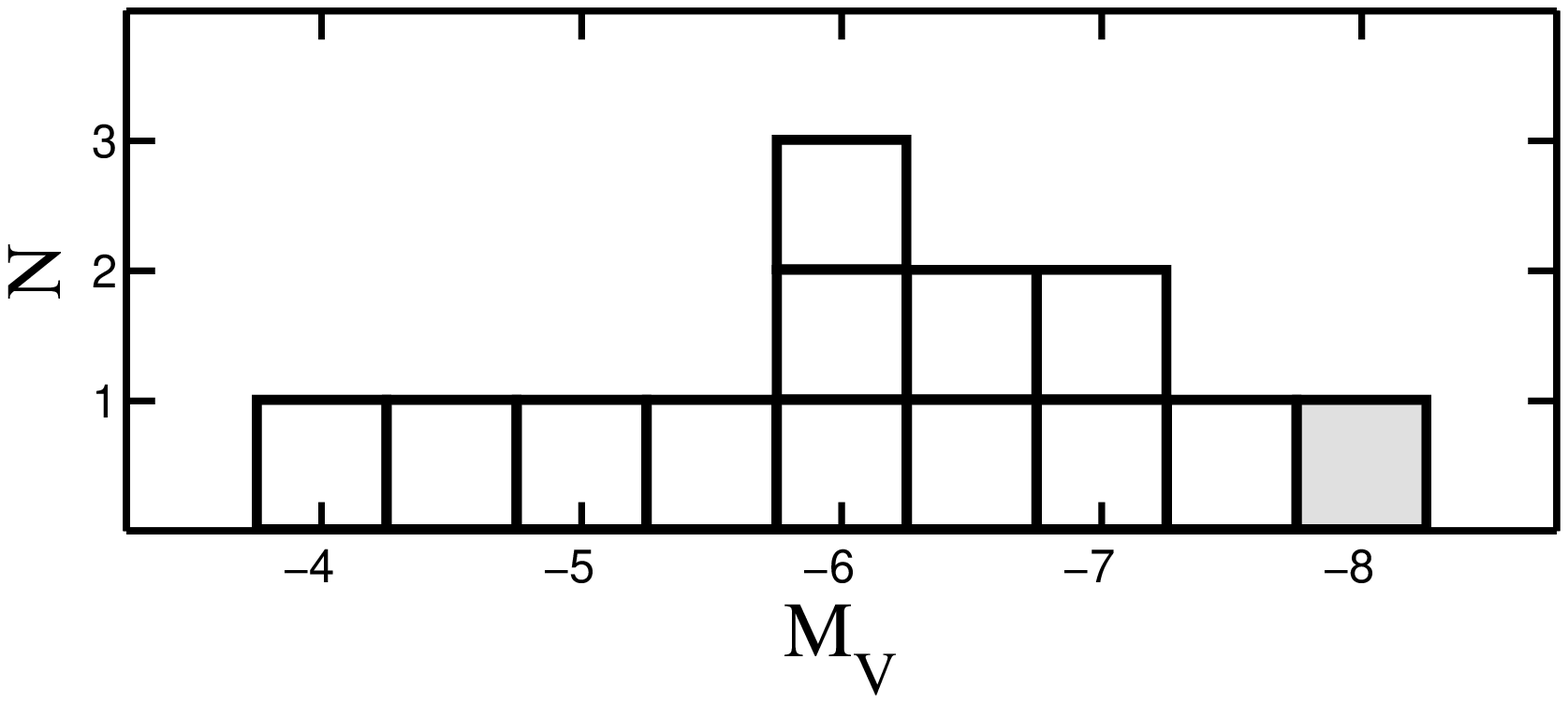}
\end{center}
\caption{}
\label{M_V}
\end{figure*}

\clearpage
\begin{figure*}[p]
\begin{center}
\hspace*{-1cm}
\includegraphics[width=10cm]{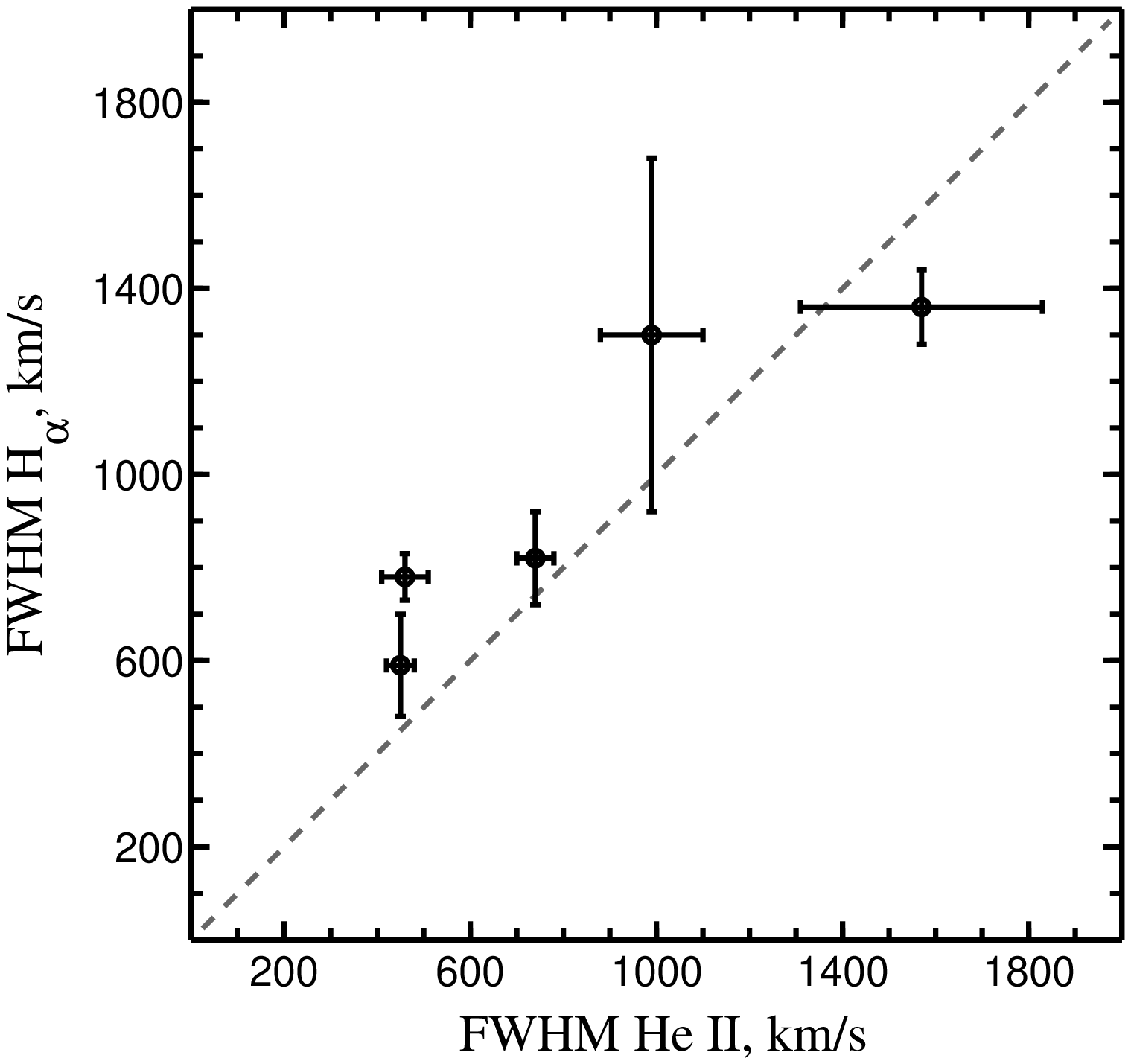}
\end{center}
\caption{} 
\label{fig3_width_he_ha}
\end{figure*}

\clearpage
   
{\large \bf Supplementary}\\
   
The Supplementary part provides details on the Subaru spectroscopic
observations, data reduction, spectral classification of the observed
spectra, comparison of the spectra with those of SS\,433, and
interpretation of the results.

\vspace{0.5cm}
\noindent
{\bf 1~~ Observations}\\

We observed ULX counterparts on 2011 February 25--28 with the Faint 
Object Camera and Spectrograph (FOCAS)$^{28}$, installed at the 
Cassegrain focus of the Subaru telescope. FOCAS was operated with the 
300B grism without filter, providing a higher efficiency with a spectral
range of 3650--8300~\AA. The range was limited by scattered-light
contamination at shorter wavelengths and by the order overlap at longer ones.
Thus, the 3800~-- 7000~\AA\ range was used. Depending on seeing conditions,
we used 0.4--0.8 arcsec slits, providing a spectral resolution of
1000~--~500, respectively. The seeing was $\approx\,$1.1 arcsec on the
first night, continuously improving to 0.4--0.6 arcsec on the last night.

Our ULX targets are very faint, V~$=$~21.5, 22.0, 23.0, and 22.5 in
Holmberg\,II, Holmberg\,IX, NGC\,4559, and NGC\,5204,
respectively$^{6,29-32}$. We tried to observe each target in each night to 
search for possible spectral and photometrical variability. With the
exception of the first night when we observed only one target (in
Holmberg\,II), the sky condition was mostly photometric. In total we
obtained the net exposure of 280, 260, 300, and 320 minutes for the
targets in Holmberg\,II, Holmberg\,IX, NGC\,4559, and NGC\,5204,
respectively.

The data reduction was performed in a standard manner. Using Th--Ar lamp,
we achieve the wavelength calibration accuracy better than 10~km s$^{-1}$ in
each spectra. However, because of the different instrumental flexure
between the arc calibration spectra, the accuracy was degrading. We
estimate a final accuracy in the night sum spectra as 15~km s$^{-1}$. 
The flux calibration is made by using standard stars observed every night.

The spectra extraction was not simple due to the nebular emission that
surrounds almost all the ULX counterparts. We extracted the spectra
with Gauss-profile aperture, and made two versions of the spectra. In
the first one, 
we used extended regions of 4--10 arcsec for the background.
Although this gives a higher S/N in the final spectra, the hydrogen 
lines originating from the nebulae may strongly contaminate the ULX
spectrum. In the second version,  
we used very small regions for the background, $\approx$1 arcsec, and
produced the spectra with the least contamination by the nebular
lines. The last version was used mainly for measurements of hydrogen lines.

The reduced spectra are given in Supplementary Fig.\,1.  For
Holmberg\,IX, we used the small regions for the background, while for
NGC\,4559 and NGC\,5204, we used the extended regions. 
For Holmberg\,II, where a very bright nebula is concentrated at the 
central part, we used an intermediate aperture for the background.  
In Fig.\,1, we show the spectra with the small apertures for the upper
three targets. The different apertures are adopted for Holmberg\,II 
and NGC\,4559 to obtain better insight how the nebulae may distort the
final spectra.

\vspace{0.5cm}
\noindent
{\bf 2 ~~Optical Spectra}\\

We produced the normalized spectra for better inspection of spectral
features (Fig.\,1). The main feature in all the spectra is a broad
He\,II\,$\lambda 4686$ emission line.  The most narrow one is found from
Holmberg\,IX, with an FWHM\,$\approx 450$\, km s$^{-1}$, and the most
broad one is from NGC\,5204, with an FWHM\,$\approx 1570$\,km s$^{-1}$
(all line widths are corrected for the spectral resolution). Note that
the relatively narrow line in Holmberg\,IX is notably broader than 
any nebular emission and it is not contaminated with nebular He\,II 
emission.

Previously, optical spectra have been obtained for M\,101\,ULX-1,
Holmberg\,IX\,X-1, NGC\,5204\,X-1, NGC\,1313\,X-2, NGC\,5408\,X-1,
NGC\,7793\,P13$^{14,16,18,33,34}$. 
In all these targets the He\,II\,$\lambda 4686$ emission has been detected. 
There was one exception, NGC\,5204\,X-1, but we confidently
detect the He\,II line from this ULX with our new data.

We conclude that all ULX counterparts ever spectrally observed have the
same feature in their spectra, that is, a broad He\,II emission line. We
also clearly detect broad H$\alpha$, H$\beta$ lines and 
He\,I\,$\lambda 6678, 5876$ lines (Supplementary Fig.\,1). There is also some
hints on the Bowen C\,III/N\,III blend (4640 - 4650~\AA). Although the 
H$\beta$ line is affected by nebular emission in spite of our careful 
extraction, its broad wings are clearly detected. It is obvious
that the emission lines are formed in stellar winds or disc winds.

Since all the spectra of the ULX counterparts are very similar to one
another, we analyze the mean parameters of the lines.
We find that the average line width of H$\alpha$ is broader than 
that of He\,II, FWHM(He\,II)/FWHM(H$\alpha$) $\approx$\,0.8.
We confirm that there is a good linear correlation between the He\,II 
and H$\alpha$ line widths in each object.
The distributions of the observed line widths are plotted in Fig.\,2. 

We find that the gas where the lines are formed is very hot, because 
EW(He\,II)/EW(H$\beta$) $\approx$\,2.2,
EW(He\,II)/EW(H$\alpha$) $\approx$\,0.36, and
EW(He\,II)/EW(He\,I$\lambda 5876$) $\gtrsim 3.6$. 
The hydrogen abundance is nearly normal in the winds, because the Pickering series are weak, 
EW(He\,II $\lambda$5411)/EW(H$\beta$) $\lesssim 0.27$ 
(Supplementary Table 1). We note that EW of the H$\beta$ line may
be uncertain because of substraction of nearby nebulae components, especially 
in NGC\,4559 and NGC\,5204. The H$\alpha$ line is measured with much 
better accuracy.

We detect a strong radial velocity variability of the He\,II\,$\lambda
4686$ line from night to night (Supplementary Fig.\,3). 
We have not enough data to study orbital variability, however, it was 
not possible to obtain dynamical mass constraints in previous spectra$^{15,16,35}$.
The He\,II line is free from contamination by the nebular He\,II emission in the 
three ULXs except in Holmberg\,II. This means that the radial velocity variability
of the line is real. The line changes not only its position but also its
intensity and width. In Supplementary Fig.\,2 we show the line variability 
in EWs and FWHMs from night to night. An accuracy depends on the He\,II 
profiles. It is less than 0.5~\AA\ (EW) and 30\,km/s (FWHM)
in the most regular lines of Holmberg\,IX\,X-1, and it is up to 1.5~\AA\ and 
250\,km/s in structured He\,II lines in NGC\,5204\,X-1 (Fig.\,1).

\vspace{0.5cm}
\noindent
{\bf 3~~ What the Spectra Resemble?}\\

Among stellar spectra, such a strong He\,II line with a nearly normal
hydrogen abundance can be found only in stars recently classified as
O2--3.5If*/WN5--7$^{9}$. We omit index * hereafter, which means a
stronger ionization as indicated by N\,IV/N\,V lines. They are the 
hottest transition stars, whose classification is based on the H$\beta$ 
profile, tracing the increasing wind density (i.e., the mass loss rate)
from O2--3.5If to O2--3.5If/WN5--7, and to WN5--7. In Supplementary Fig.\,2, 
we show the classification diagram of WN stars$^{7}$ 
for LMC and Galactic objects. We supplement the diagram with additional
stars recently classified$^{9}$. The diagram plots stars in
accordance with their wind velocity (FWHM) versus the photosphere 
temperature and mass loss rate (EW).
Three known LBV transitions (LBV\,--\,WNL) between their hot and cool
states in AG\,Car, V\,532 in M\,33, and HD\,5980 in SMC are also shown in
the figure. Consequent states in each LBV transition are connected with
the lines. In their hotter state where the He\,II line becomes stronger, 
the LBVs fit well the classical WNL stars$^{8}$.

We see that the ULX counterparts occupy a region at the He\,II diagram between
O2--3.5If and WN5--7 (Supplementary Fig.\,2). This is also a region of 
``intermediate temperature LBV'' V\,532 and the ``LBV excursion'' of 
HD\,5980. However, their behavior in the diagram is nothing like stars. 
The variability of the He\,II lines 
of our counterparts in three consequent nights is shown by the points 
connected by the lines. If the ULX counterpart spectra were produced from 
donor stars, variable surface gravity at about the same phothospheric 
temperature would be required. Instead, the spectra may be
formed in unstable and variable winds formed in accretion discs. This
idea agrees with the fact that we do not find any regularities between
the EW, FWHM, and radial velocity of the He\,II line.

In the figure, we also present two recently discovered extragalactic
black holes NGC\,300\,X-1$^{36}$ and IC\,10\,X-1$^{37}$ together with the 
soft ULX transient M\,101\,ULX-1. The black holes in NGC\,300\,X-1 and 
IC\,10 have luminosities $L_{\rm X} \sim 3 \times 10^{38}$\,erg s$^{-1}$, 
about the same as that of Cyg\,X-3, which certainly contains a WN-type donor
star$^{38}$. 

The same may be proposed
for the transient source M\,101\,ULX-1 on the basis of its location in the 
diagram. It has been recently found that this source indeed contains a WN8 type
donor$^{18}$, although its orbital period is $\sim 40$ times longer than
in Cyg\,X-3 and $\sim 6$ times longer than in two other WR X-ray 
binaries. This may be the reason for the difference of M\,101\,ULX-1 
from the other WR X-ray binaries, i.e., its transient nature.
Because of the large size of the Roche lobe around the compact star, the 
accretion disc is likely to be formed with a partial ionization zone$^{18}$. 
In steady states, M\,101\,ULX-1 has a luminosity of an order of magnitude
smaller than that of the Cyg\,X-3-type objects, although in an outburst peak 
the luminosity becomes an order of magnitude larger, close to the low
boundary of the ULXs luminosity range. 

The X-ray luminosities comparable with that of Cyg\,X-3, which are in line 
with the wind accretion and the short orbital periods, the location in 
the He\,II diagram, 
the detail analysis of optical spectra of the WR X-ray binaries$^{36,37,18}$ 
confirm that their optical spectra come from WNL donors. 

The only known supercritical accretor in our Galaxy SS\,433 seems to be
the most close relative to the ULXs. It has been proposed$^{3}$ that
SS\,433 supercritical disc's funnel will appear as an extremely bright
X-ray source when observed in a nearly face-on geometry.
The intrinsic luminosity of SS\,433 (mainly in UV) is 
$\sim 10^{40}$\,erg s$^{-1}$ and all this budget is formed in the 
disc$^{3}$. Nevertheless, the X-ray luminosity of SS\,433
is not high, $\sim 10^{36}$\,ers s$^{-1}$, because the emission is 
blocked by the thick accretion disc observed nearly edge-on.  
If one has a chance to
observe the whole funnel, one may discover all the bolometric radiation
in X-rays. Apparent X-ray luminosity of the face-on SS\,433 is expected
to be even larger than its bolometric luminosity, because of geometrical 
collimation in the funnel with a factor of 
$B = 2\pi / \Omega_f \sim 3-5$, where $\Omega_f$ is the solid angle of 
the funnel. The existence of SS\,433 demands forthcoming X-ray
sources with luminosities of a few $\times 10^{40}$\,erg s$^{-1}$.

SS\,433 in Supplementary Fig.\,2 diagram nearly meets the ULX region.
However, the ULX
winds are notably hotter than the SS\,433 wind. Comparing the ULX
spectra with that of SS\,433$^{10}$ taken with the same
instrument, we find that the average EWs of H$\alpha$, H$\beta$, and He\,I
$\lambda 5876$ lines in the ULXs are exactly $\approx 12$ times smaller than those
in SS\,433, whereas the EWs of two He\,II lines ($\lambda \lambda 4686,
5412$) are only $2.2$ times smaller than in SS\,433 (Supplementary Table\,1). 
We conclude that 
the ULX winds are not such powerful, but notably hotter, than that in 
SS\,433. That is because their EWs are smaller and their He\,II/He\,I 
ratios are bigger than that in SS\,433. In this sense, SS\,433 remains 
the unique object. Indeed, the presence of the persistent jets in 
SS\,433 may be explained by its extremely large mass accretion rate. 

Below we discuss possible interpretations of the ULX optical spectra: 
a donor star and a supercritical discs with a stellar-mass black hole.

\vspace{0.5cm}
\noindent
{\bf 4 ~~Are the ULX Spectra of the Donor Stars?}\\

If the optical spectrum is formed in the donor star of WNL type, the
He\,II line widths indicate the stellar wind velocity. We adopt this
value as $V_{\rm W} = 900 \,v_{900}$\,km s$^{-1}$, where $v_{900}$ is the wind
velocity normalized by 900\,km s$^{-1}$. The wind must supply a mass accretion
rate $\dot M_{\rm X}$ to provide the observed X-ray luminosity of
$L_{\rm X} \sim 10^{40}$\,ers s$^{-1}$. If we adopt an accretion
efficiency $\eta = L_{\rm X}/\dot M_{\rm X} c^2 = 0.1$, 
we need $\dot M_{\rm X} \sim 1.8 \times 10^{-6}~\msun$ y$^{-1}$. 
Note this is the super-Eddington rate for black holes with masses 
smaller than 60~${\msun}$. Both the mass accretion rate
and black hole mass must be regarded as the lower limits because of 
advection of heat and radiation$^{27}$.

In the close binary consisting of a black hole and a WN star, the
donor's mass loss rate in the wind must be $\dot M_{\rm WN} = \dot
M_{\rm X} 4 \pi a^2 / \pi R_{\rm c}^2$, where $a$ is the binary
separation, $R_{\rm c} \approx G M_{\rm BH} / (V_{\rm W}^2 + V_{\rm
orb}^2)^2$ is the capture (Bondi) radius and $V_{\rm orb}$ is the
relative orbital velocity of the companion. Adopting the mass ratio in
the binary as $q = M_{\rm WN}/M_{\rm BH} = 2$ and using Kepler's law, we
find $\dot M_{\rm WN} \approx 300 \dot M_{\rm X} (p_1/m_{10})^{4/3}
v_{900}^4 \approx 5 \times 10^{-4}$~${\msun}$ y$^{-1}$, where $p_1$ and
$m_{10}$ are the orbital period and black hole mass 
in units of 1 day and 10~$\msun$, respectively. Even for such a short
period the donor's mass loss is too large for WNL stars$^{39}$. 
It is because the wind--fed accretion is not effective. The
orbital velocity of the donor in the binary will be $\sim 220$~km
s$^{-1}$. If one assumes smaller orbital periods or larger black hole
masses, one finds higher orbital velocities, which are not observed in
our observations (Supplementary Fig.\,3).

There is an exotic case of the IMBH\,+\,WN donor binary ($M_{\rm BH} \gg
10~{\msun}$, $V_{\rm orb} \gg V_{\rm W}$), where practically all the
donor's wind is accreted by the black hole and hence $\dot M_{\rm WN} / \dot
M_{\rm X} = 4(1+q)^2$. Such a system can provide the observed X-ray
luminosity of $\sim 10^{40}$\,ers s$^{-1}$. However, the donor's orbital
velocity variations must be much larger than observed value, $V_{\rm orb} \sim 2100
(m_{1000}/p_1)^{1/3}$\,km s$^{-1}$. We thus conclude that a WN donor cannot
provide the ULX-like luminosity with a black hole of any masses. Indeed,
three known examples of the short period binaries with WN donors,
Cyg\,X-3, IC\,10\,X-1, and NGC\,300\,X-1, are not ULXs, which are
$\sim$30 times fainter in X-rays.

One finds a strong argument that the He\,II emission in the ULXs cannot 
be formed on a donor star. It comes from high-quality optical spectra of
NGC\,7793\,P13, where the donor temperature of $11000 \pm 1000$\,K has 
been measured$^{14}$. 
We have studied the P13 spectra and confirm that neither the emission 
line EW nor the absorption spectrum of the B9Ia supergiant companion 
depends on the X-ray luminosity or the orbital phase of the system.
We suggest that this object is somewhat different from the other ULXs, 
and hence additional studies are needed to completely understand its nature.
Note that P13 is not a persistent ULX, whose X-ray luminosity is variable 
by a factor of 100, although it could be related to the source 
occultation by the disc.

\vspace{0.5cm}
\noindent
{\bf 5 ~~Supercritical Discs with Stellar-Mass Black Holes}\\

We have shown that the He\,II line is most likely formed in super-Eddington
accretion discs with stellar-mass black holes. 
The supercritical regime has
been first described by Shakura and Sunyaev$^{26}$. They introduced a
``spherization radius'' in the disc, $R_{\rm sp} \propto \dot M_0$. 
The disc is standard at $r > R_{\rm sp}$, 
while it becomes 
a supercritical one at all radii smaler than this point.
The SCAD is geometrically thick with strong mass loss.
The local mass accretion rate at $r < R_{\rm sp}$ is Eddington limited, 
$\dot M(r) \sim \dot M_0 r/R_{\rm sp}$.
As the result, a strong disc wind inevitably appears in the
supercritical region$^{5}$. In recent 2D RHD simulations$^{27}$, 
which take into account both heat advection and photon
trapping, the main ideas of the Shakura--Sunyaev's SCAD approach have
been confirmed.

In standard accretion discs, the bolometric luminosity is scaled with
the mass accretion rate as $L \propto \dot M_0$, whereas, in SCADs, the
luminosity depends on black hole mass and is nearly independent of mass
accretion rate as $L \propto L_{\rm Edd} (1 + a \ln(\dot M_0 / \dot
M_{\rm Edd}))$ $\propto M_{\rm BH}$\,$^{26}$, where $L_{\rm Edd}$ and $\dot
M_{\rm Edd}$ are the Eddington limit and corresponding mass accretion
rate, respectively, and $a \sim 0.5-0.7$ is the parameter accounting for
advection$^{5}$. There are two more differences in SCADs from standard
discs, (i) the X-ray radiation is geometrically collimated ($2\pi /
\Omega_f$) in the wind funnel, and (ii) the UV/optical luminosity
does depend stronger on the mass accretion rate than it is in a standard 
disc. In the SCAD all the gas in excess of the critical accretion rate 
is expelled as a disc wind$^{13}$, which reprocesses the disc X-ray 
radiation. The wind forms a funnel whose size is nearly proportional
to the mass accretion rate. This issue may resolve the dimness of 
the ULX optical counterparts with respect to SS\,433 (Fig.\,2).

The wind velocity $V$ is expected to be virial at the spherization 
radius$^{26}$, $V \propto M_{\rm BH}^{1/2} R_{\rm sp}^{-1/2}$. The wind 
photosphere radius is given as $R_{\rm ph} \propto \dot M_0 V^{-1} \propto \dot
M_0^{3/2} M_{\rm BH}^{-1/2}$\,$^{3}$. The SCAD luminosity, $L \propto
M_{\rm BH} \propto R_{\rm ph}^2 T^4$, is partially radiated at the
wind photosphere. Combining these dependences, we find the photosphere
temperature dependence as $T_{\rm ph} \propto \dot M_0^{-3/4} M_{\rm BH}^{1/2}$. 
In the optical Rayleigh--Jeans region, the luminosity is expressed as 
$L_{\rm V} \propto R_{\rm ph}^2 T \propto \dot M_0^{9/4} M_{\rm BH}^{-1/2}$. 
Hence, the optical luminosity of the SCAD may strongly depend on the 
mass accretion rate.

In the picture described by Shakura-Sunyaev$^{26}$ the excess gas is 
ejected from the spherization radius $R_{\rm sp}$ at a virial velocity. 
This scale is estimated as $R_{\rm sp} \sim \frac{\kappa \dot M_0}{8 \pi c}$,
where $\kappa = 0.34\,\,\text{cm}^2\,\text{g}^{-1}$ is the Thomson's 
opacity. In SS\,433 one finds $R_{\rm sp} \sim 3 \times 10^9$\,cm, and 
the corresponding virial velocity is $V \sim 9400$\,km s$^{-1}$. The 
velocity is one magnitude bigger than that observed in SS\,433 and in
the ULX counterparts. It is possible, however, that the actual situation 
is much more complex than the simpified picture that the gas is directly 
ejected without any interactions in the wind.
To account for the discrepancy between the predicted velocity and
observed ones, it is suggested, in modelling of the SS\,433
wind$^{40}$, that the escaping gas is mixing because of effective mass
loading of the funnel region. Also, it is known from hydrodynamic
simulations$^{41,27}$ that some portions of the gas are launched at velocities
less than the virial one and return to the disc, but the excess gas
eventually leaves the disc.

Note that the gas ejected at $R_{\rm sp}$ is optically thick in
continuum.  The process occurs very deeply under the photosphere,
$R_{\rm ph}$, whose size is about three orders of magnitude larger than
the launching radius.  At $R > R_{\rm ph}$, the gas remains to be
optically thick in lines. That is a principal difference between
supercritical disc outflows and WNL/LBV winds; in supercritical discs,
the gas is launched in very inner regions. Above the photosphere,
however, the winds may behave about the same way.

We cannot know all acceleration processes in disc winds without
detailed modelling up to the photosphere size. If we assume that the
gas velocity is constant, we find $R_{\rm ph} \propto \dot M_0$, $T_{\rm
ph} \propto \dot M_0^{-1/2} M_{\rm BH}^{1/4}$, and $L_{\rm V} \propto
\dot M_0^{3/2} M_{\rm BH}^{1/4}$. Indeed, estimates of these values are
based on the velocity around the photosphere, where the gas velocity
cannot change rapidly. In such a case, we again see that the optical
luminosity of the winds does depend notably on the gas accretion rate.

At supercritical accretion rates ($\dot M_0 > \dot M_{\rm Edd}$), the
extended photosphere of the wind hides its formation region, and hence
$R_{\rm ph} > R_{\rm sp}$, where $R_{\rm ph}$ is the location of the
photosphere. It is known that in SS\,433, $\dot M_0 \gg \dot M_{\rm
Edd}$\,$^{3}$. It is the brightest object in the $V$-band among all 
studied ULX counterparts (Fig.\,2), by a factor from 2 to 60. Using the
scaling relations derived above {\bf ($L_{\rm V} \propto \dot M_0^{9/4}$)}, 
we find that the mass accretion rates
in the ULXs are smaller than that in SS\,433 by a factor from 1.5 to 6 and
their wind temperatures are higher by 1.4--4. SS\,433 wind
has a temperature of $\sim 50$\,kK$^{3}$ in its precession phase where the
disc is the most open to the observer. We thus estimate the wind
temperatures in the ULXs to be 70--200\,kK. (If we take another 
scaling relation, $L_{\rm V} \propto \dot M_0^{3/2}$, we find that 
in the ULXs the mass accretion rates are smaller by a factor of 1.6~--~15,
while the wind temperatures are higher by the same factor, 1.4--4,
compared with SS\,433).

The ULX winds are very hot, because they are directly heated by the strong
X-ray radiation. Thus, the simple SCAD model may explain both the relative
dimness of the ULX counterparts in the optical bands
as well as the higher ionization stage of the winds than in SS\,433.
The He\,II and H$\alpha$ lines are produced in outer parts of the wind,
not in the wind launching region where the gas is too hot to produce
optical lines.

Recent observations of broad band X-ray spectra of ULXs$^{42, 43, 44}$
reveal that they have spectral cutoffs at relatively low energies
(around ten keV), which are not seen in the spectra of sub-Eddington 
black hole binaries in the low/hard or high/soft states$^{45}$. 
Thus, the physical conditions of the innermost disc are indeed quite 
different between ULXs and sub-Eddington black hole binaries.
The spectral cutoff in ULXs may be interpretated
as indication of the supercritical accretion$^{46}$ 
(the "ultraluminous state``), where a massive a wind may completely 
envelope the inner-disc regions, creating a cool Comptonizing 
photosphere. 
The broadband X-ray observations together with the optical observations 
presented here strongly support the picture for ULXs as 
supercritical discs.

One of the appreciable properties of some ULXs is quasi-periodic
oscillations (QPOs) with periods of typically tens of seconds$^{1}$. 
The QPOs are stronger in hard X-ray bands.
Remarkable modulations of the disc luminosity and the accretion rate 
through the inner edge of the disc were found$^{47}$ in time-dependent 
two-dimensional radiation hydrodynamical calculations performed for a 
supercritical accretion disc by adopting the parameters of SS\,433.
These modulations produce recurrent hot blobs in the SCAD funnel
to be become apparent as high amplitude QPOs. A time scale of the 
QPOs$^{47}$ is in about the same range as those observed in the ULXs.

Thus, we interpret that all these ULXs with X-ray luminosities of 
$L_{\rm x} \sim 10^{40}$\,erg~s$^{-1}$ constitute a homogeneous class of 
stellar-mass black holes with supercritical discs. They are distinct 
population from hyperluminous X-ray sources (HLXs), which are most 
probably IMBHs$^{48}$.

\clearpage
\noindent
{\bf Supplementary Figure 1: Calibrated spectra of the ULX optical counterparts. }
From top to bottom: the ULX in Holmberg\,II, Holmberg\,IX, NGC\,4559,
and NGC\,5204. The two upper spectra were obtained on February 28, while
the rest are the summed spectra from the three nights. For better
visualization we add flux offsets of 1.8, 1.2 and 0.6
($10^{-17}$\,erg/cm$^2$s\,\AA~) for the Holmberg\,II, Holmberg\,IX, and
NGC\,4559 ULXs, respectively. Besides obviuos hydrogen lines 
we mark He\,II lines ($\lambda 4686$ and $\lambda 5412$) and
He\,I lines ($\lambda 5876$ and $\lambda 6678$), thick bar indicates position of 
the Bowen blend C\,III/N\,III $\lambda \lambda 4640 - 4650$.\\ \\

\noindent
{\bf Supplementary Figure 2: Classification diagram of WNL stars in the 
LMC and our Galaxy$^{7}$. }
The black open squares, triangles, and circles mark
WN\,8, WN\,9--10m, and WN\,11 stars, respectively.
The blue filled circle denotes $\zeta$\,Pup. Other Galactic and LMC
stars$^{9}$ are O2If and O3If (open blue circles), O2If/WN5, O2.5If/WN6,
O3If/WN6, and O3.5If/WN7 (blue crosses), and WN6ha and WN7ha stars (open blue
squares). There are three known LBV\,--\,WNL transitions (AG\,Car, V\,532,
and HD\,5980) in this diagram$^{8}$. Consequent states of each 
LBV star are connected with the lines.
Positions of our four ULX counterparts are also shown (connected with 
lines to show variability from night to night), together with those of
SS\,433, NGC\,7793\,P13, NGC\,5408\,X-1, NGC\,1313\,X-2, M\,101\,ULX-1, NGC\,300\,X-1, 
and IC\,10\,X-1$^{10,14,15,16,18,36,37}$. \\

\noindent
{\bf Supplementary Figure 3: Radial velocities of He\,II (blue) and
H$\alpha$ (red) lines for three consequent nights.} From top to bottom:
the ULX in NGC\,4559, NGC\,5204, and Holmberg\,IX. For the last 
target we show only the He\,II velocities (shifted by $-300$\,km
s$^{-1}$ from those of H$\alpha$). 
To estimate the error in the radial velocity, we produced simulated
spectra of the He\,II (or H$\alpha$) line with the same profile
(modelled by a Gaussian) and random noises as those of the real
data. We regard the scatter ($1\sigma$) of their best-fit radial
velocities as the error, which is attached to each data point.\\ \\

\noindent
{\bf Supplementary Table 1: Mean emission line equivalent widths} (in Angstroms) and their ratios 
of the main emission lines of our ULXs sample and SS\,433. The data of SS~433 
are taken from the literature$^{10,24}$. The last column presents 
the ULX equivalent widths and the line ratios with respect to those 
in SS\,433.

\clearpage 
\renewcommand{\figurename}{Supplementary Figure 1}
\renewcommand{\thefigure}{}
\begin{figure*}[p]
\hspace*{-1.6cm}
\includegraphics[width=14cm]{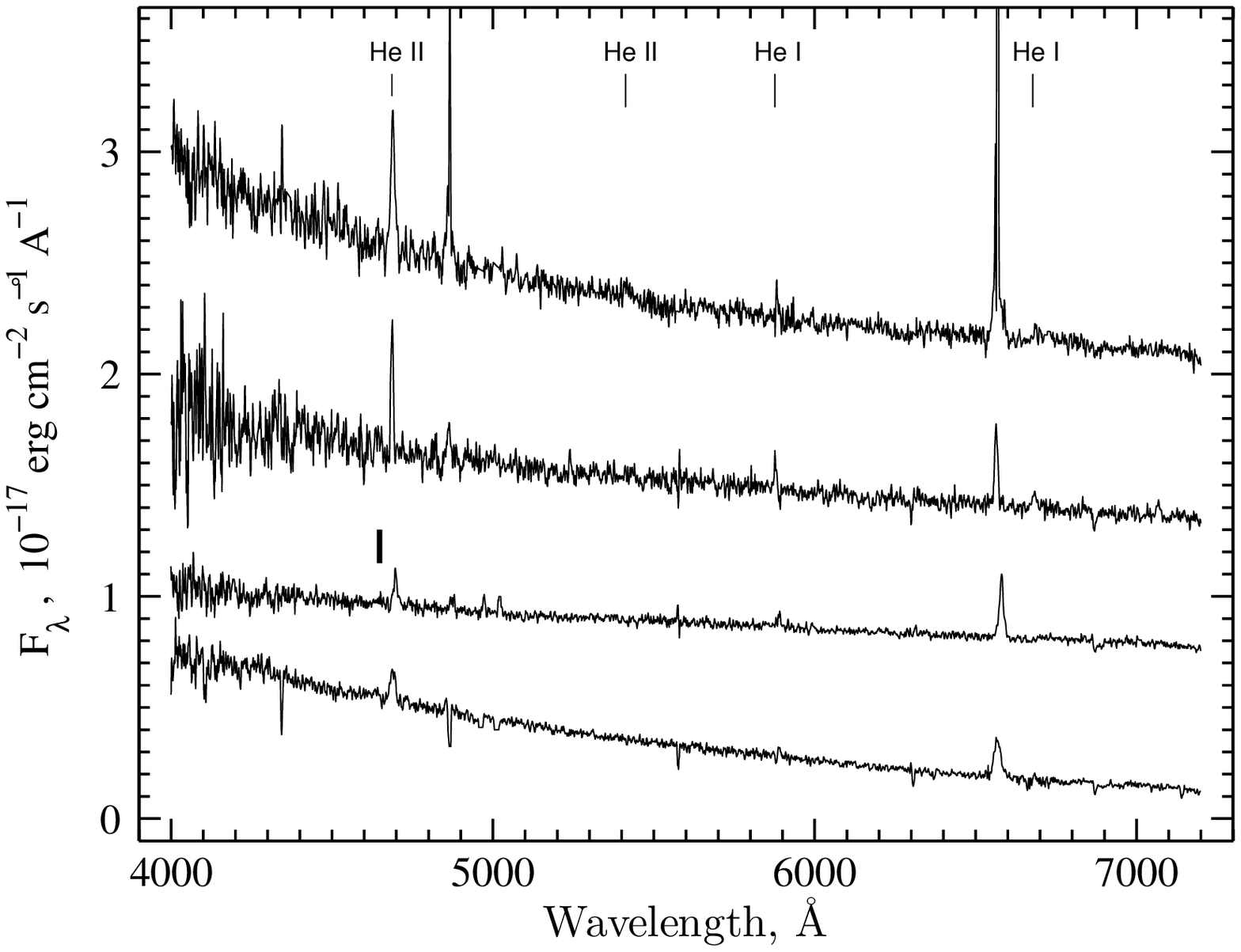}
\caption{} 
\label{whole_spectra}
\end{figure*}

\clearpage
\renewcommand{\figurename}{Supplementary Figure 2}
\renewcommand{\thefigure}{}
\begin{figure*}[p]
\hspace*{-2.1cm}
\includegraphics[width=15cm]{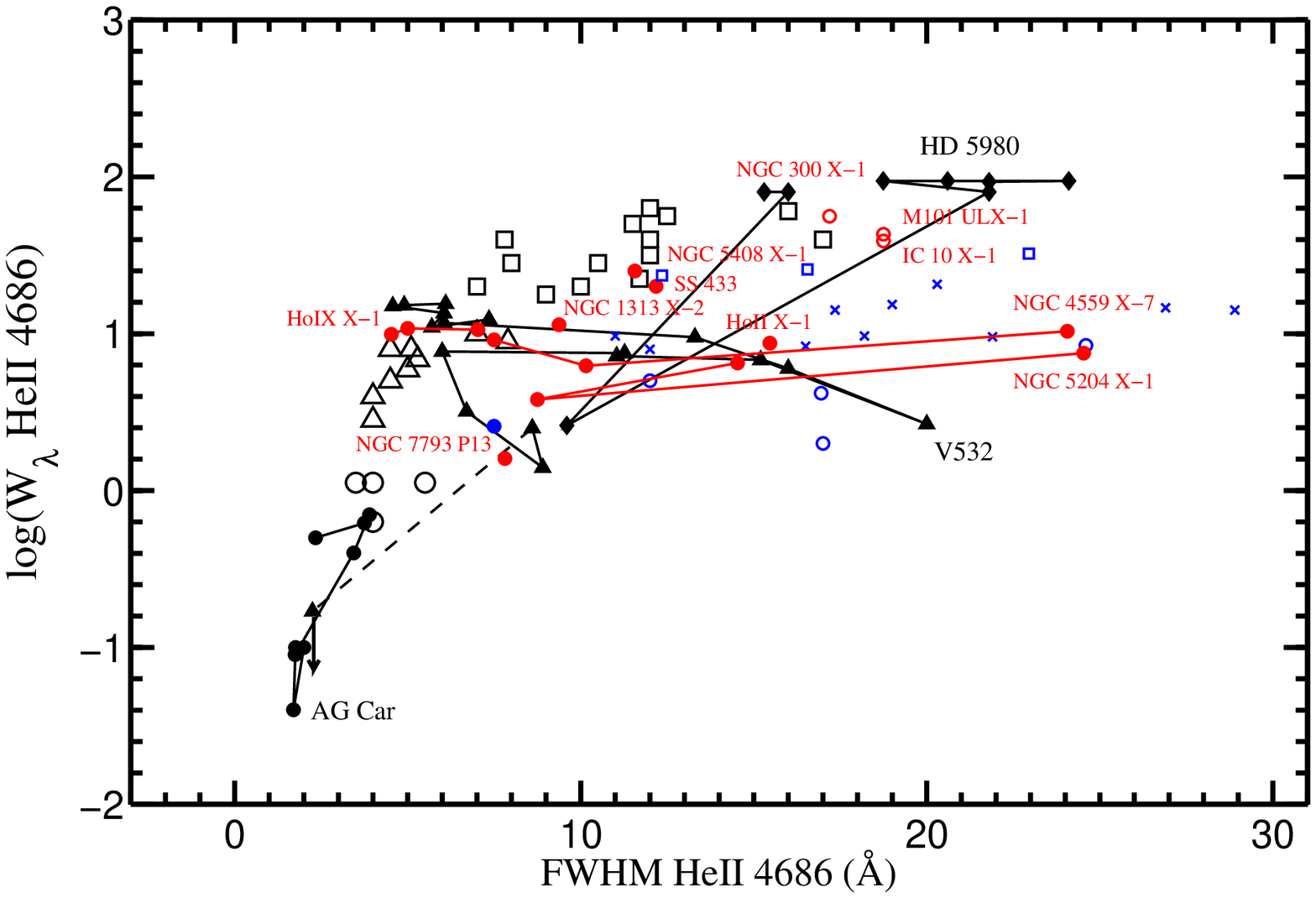}
\caption{}
\label{fig2_HeII_diagram}
\end{figure*}

\clearpage
\renewcommand{\figurename}{Supplementary Figure 3}
\renewcommand{\thefigure}{}
\begin{figure*}[p]
\hspace*{0.2cm}
\includegraphics[width=10cm]{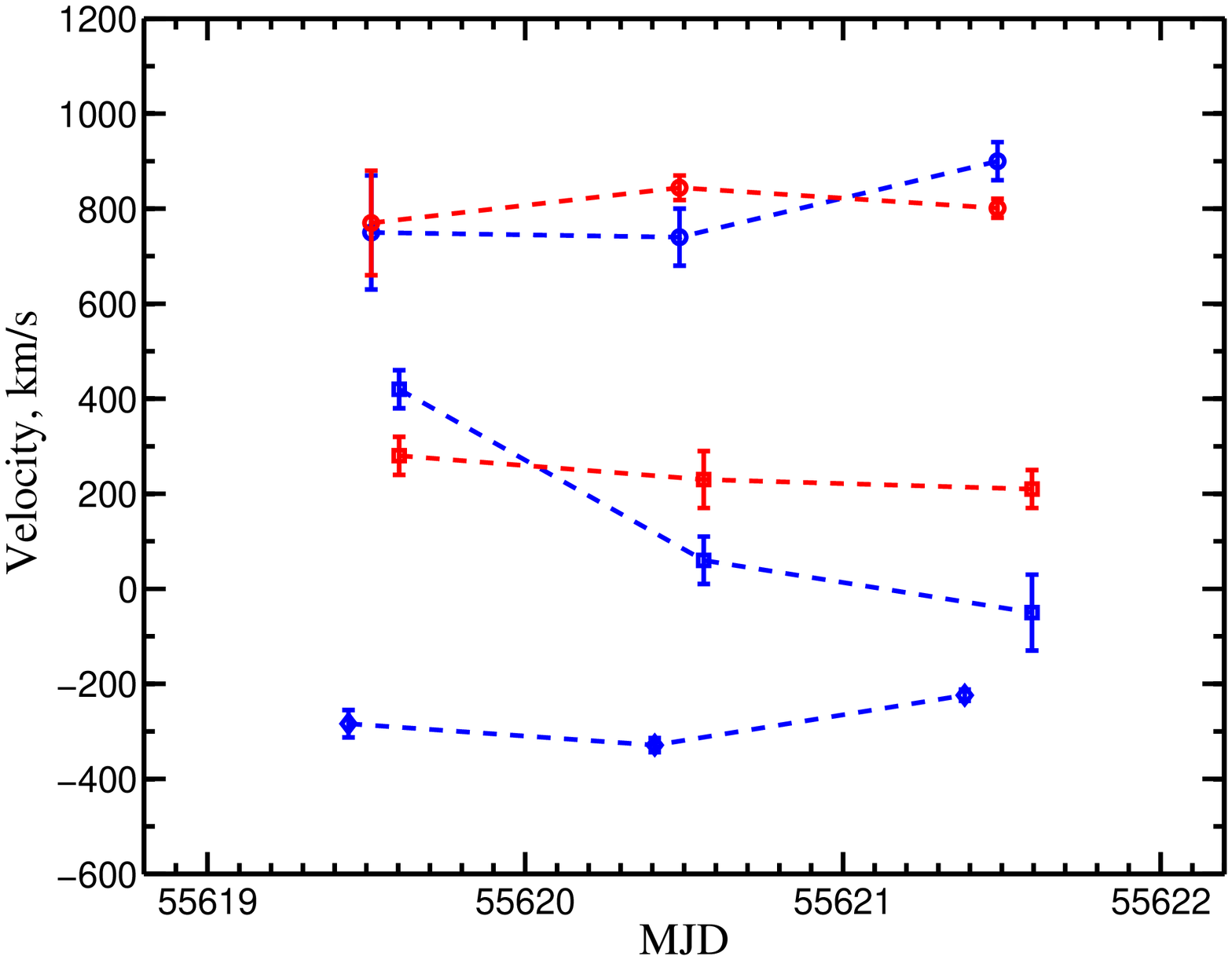}
\caption{}
\label{fig3_radvel}
\end{figure*}

\clearpage

\renewcommand{\tablename}{Supplementary Table 1} 
\renewcommand{\thetable}{}
\begin{table*}[!tp]
\caption{}
\label{measurements}
\bigskip
\hspace*{0.9cm}
\begin{tabular}{lcccc}
\hline
&&&&\\
& ~~EW$_{\rm ULX}$~~ & ~~EW$_{\rm SS\,433}$~~ &  ULX\,/\,SS\,433 \\
&&&&\\
\hline
&&&&\\
He\,II $\lambda$4686       & 9.0$\pm$0.6            & 20       & 0.45$\pm$0.03\\
He\,II~$\lambda$5412       & $\lesssim$ 1.1$\pm$0.3 & 2.5      & $\lesssim$ 0.44$\pm$0.12\\
He\,I~$\lambda$5876        & $\lesssim$ 2.5$\pm$0.7 & 24.5     & $\lesssim$ 0.10$\pm$0.03\\
$\Hb$                      & 4.1$\pm$1.2            & 50       & 0.082$\pm$0.024\\
$\Ha$                      & 25$\pm$4               &320$\pm$20& 0.078$\pm$0.012\\
$\Hb/\Ha$                  & 0.16$\pm$0.06          & 0.16     & 1.0$\pm$0.4\\
He\,II~$\lambda$4686/$\Ha$ & 0.36$\pm$0.06          & 0.06     & 6.0$\pm$1.0\\
He\,I~$\lambda$5876/$\Ha$  &$\lesssim$ 0.10$\pm$0.03& 0.08     & $\lesssim$1.3$\pm$0.4\\
&&&&\\
\hline
\end{tabular}
\end{table*}

\end{document}